\documentclass[12 pt,twoside,aps,prd,amsmath,amssymb,
tightenlines,showpacs,showkeys,eqsecnum]{revtex4-1}

\usepackage{graphicx}
\usepackage{mathptmx}
\usepackage{bm}
\newcommand {\pr}{\partial}
\newcommand {\cG}{\cal G}
\newcommand {\cD}{\cal D}
\newcommand {\bg}{\bar \gamma}
\newcommand {\G}{\Gamma}
\newcommand {\bp}{\bar \psi}
\newcommand {\p}{\psi}

\def\myfigure #1#2#3#4
{\begin{figure}[ht]\begin{center}
\includegraphics[width=#2 \textwidth]{#1.eps}
\parbox[t]{#4\textwidth}{\caption{#3}\label{#1}}
\end{center}\end{figure}}

\def \myfigures #1#2#3#4#5#6#7#8
{\begin{figure}[ht]
    \begin{center}
        \includegraphics[width=#2 \textwidth]{#1.eps}
        \hfill
        \includegraphics[width=#6 \textwidth]{#5.eps}
        \parbox[t]{#4\textwidth}{\caption {#3}\label{#1}}
        \hfill
        \parbox[t]{#8\textwidth}{\caption {#7}\label{#5}}
    \end{center}
\end{figure} }

\begin{document}
\baselineskip -24pt
\title{Nonlinear Spinor Fields in Bianchi type-$VI_0$ spacetime}
\author{Bijan Saha}
\affiliation{Laboratory of Information Technologies\\
Joint Institute for Nuclear Research\\
141980 Dubna, Moscow region, Russia} \email{bijan@jinr.ru}
\homepage{http://bijansaha.narod.ru}

\begin{abstract}

Within the scope of Bianchi type-$VI_0$ spacetime we study the role
of spinor field on the evolution of the Universe. It is found that
the presence of nontrivial non-diagonal components of
energy-momentum tensor of the spinor field plays vital role on the
evolution of the Universe. As a result of their mutual influence
there occur two different scenarios. In one case the invariants
constructed from the bilinear forms of the spinor field become
trivial, thus giving rise to a massless and linear spinor field
Lagrangian. According to the second scenario massive and nonlinear
terms do not vanish and depending on the sign of coupling constants
we have either an expanding mode of expansion or the one that after
obtaining some maximum value contracts and ends in big crunch
generating spacetime singularity. This result shows that the spinor
field is highly sensitive to the gravitational one.

\end{abstract}



\maketitle

\bigskip

\section{Introduction}

Thanks to its flexibility to simulate the different characteristics
of matter from perfect fluid to dark energy and its ability to
describe the different stages of the evolution of the Universe,
spinor field has become quite popular among the cosmologists
\cite{henneaux,ochs,saha1997a,saha1997b,saha2001a,greene,saha2004a,
saha2004b,ribas,saha2006c,saha2006d,saha2006e,saha2007,
souza,PopPLB,FabJMP,ELKO,PopPRD,PopGREG,FabIJTP,kremer} . But some
recent study \cite{sahaIJTP2014,sahaAPSS2015} suggests that flexible
though it is, the existence of non-diagonal components of the
energy-momentum tensor of the spinor field imposes very severe
restrictions on the geometry of the Universe as well as on the
spinor field, thus justifying our previous claim that spinor field
is very sensitive to the gravitational one \cite{sahashikinCJP}.

In some recent papers \cite{sahaIJTP2014,sahaAPSS2015} within the
scope of Bianchi type-I cosmological model the role of spinor field
in the evolution of the Universe has been studied. It is found that
due to the spinor affine connections the energy momentum tensor of
the spinor field becomes non-diagonal, whereas the Einstein tensor
is diagonal. This non-triviality of non-diagonal components of the
energy-momentum tensor imposes some severe restrictions either on
the spinor field or on the metric functions or on both of them. In
case if the restrictions are imposed on the components of spinor
field only, it becomes massless and invariants constructed from
bilinear spinor forms also become trivial. Imposing restriction
wholly on metric functions one obtains FRW model, while if the
restrictions are imposed both on metric functions and spinor field
components, the initially BI model becomes locally rotationally
symmetric. These results motivated us to consider the other Bianchi
models and study the influence of spacetime geometry on the spinor
field and vice versa.

A Bianchi type-$VI_0$ model describes an anisotropic spacetime and
generates particular interest among physicists. Weaver
\cite{Weaver}, Ib$\acute{a}\tilde{n}$ez et al. \cite{Hoogen},
Socorro and Medina \cite{Socorro}, and Bali et al. \cite{Bali} have
studied B-$VI_{0}$ spacetime in connection with massive strings.
Recently, Belinchon \cite{Belinchon} studied several cosmological
models with B-$VI_{0}$ \& B-III symmetries under the self similar
approach. A spinor description of dark energy within the scope of a
B-$VI_0$ model was given in \cite{Saha2012}. Bianchi type $VI_{0}$
spacetime filled with dark energy was investigated in
\cite{sahaECAADE}.

In this paper we study the self-consistent system of nonlinear
spinor field and gravitational one given by the Bianchi type
$VI_{0}$ spacetime in order to clarify the role of non-diagonal
components of the energy-momentum tensor of spinor field in the
evolution of the Universe.

\section{Basic equation}

Let us consider the case when the anisotropic space-time is filled
with nonlinear spinor field. The corresponding action can be given
by
\begin{equation}
{\cal S}(g; \psi, \bp) = \int\, L \sqrt{-g} d\Omega \label{action}
\end{equation}
with
\begin{equation}
L= L_{\rm g} + L_{\rm sp}. \label{lag}
\end{equation}
Here $L_{\rm g}$ corresponds to the gravitational field
\begin{equation}
L_{\rm g} = \frac{R}{2\kappa}, \label{lgrav}
\end{equation}
where $R$ is the scalar curvature, $\kappa = 8 \pi G$, with G being
Newton's gravitational constant and $L_{\rm sp}$ is the spinor field
Lagrangian.

\subsection{Gravitational field}

The gravitational field in our case is given by a Bianchi
type-$VI_0$ anisotropic space time:

\begin{equation}
ds^2 = dt^2 - a_1^2 e^{-2mx_3} dx_1^2 - a_2^2 e^{2mx_3} dx_2^2 -
a_3^2 dx_3^2, \label{bvi0}
\end{equation}
with $a_1,\,a_2$ and $a_3$ being the functions of time only and $m$
is some arbitrary constant.

The nontrivial Christoffel symbols for \eqref{bvi0} are
\begin{eqnarray}
\G_{01}^{1} &=& \frac{\dot{a_1}}{a_1},\quad \G_{02}^{2} =
\frac{\dot{a_2}}{a_2},\quad
\G_{03}^{3} = \frac{\dot{a_3}}{a_3}, \nonumber\\
\G_{11}^{0} &=& a_1 \dot{a_1} e^{-2mx_3},\quad \G_{22}^{0} = a_2
\dot{a_2} e^{2mx_3},\quad
\G_{33}^{0} = a_3 \dot{a_3},\label{Chrysvi}\\
\G_{31}^{1} &=& -m,\quad \G_{32}^{2} = m,\quad \G_{11}^{3} = \frac{m
a_1^2}{a_3^2} e^{-2mx_3},\quad \G_{22}^{3} = -\frac{m a_2^2}{a_3^2}
e^{2mx_3}. \nonumber
\end{eqnarray}

The nonzero components of the Einstein tensor corresponding to the
metric \eqref{bvi0} are
\begin{subequations}
\label{ET}
\begin{eqnarray}
G_1^1 &=&  -\frac{\ddot a_2}{a_2} - \frac{\ddot a_3}{a_3} -
\frac{\dot a_2}{a_2}\frac{\dot a_3}{a_3} + \frac{m^2}{a_3^2}, \label{ET11}\\
G_2^2 &=&  -\frac{\ddot a_3}{a_3} - \frac{\ddot a_1}{a_1} -
\frac{\dot a_3}{a_3}\frac{\dot a_1}{a_1} + \frac{m^2}{a_3^2}, \label{ET22}\\
G_3^3 &=&  -\frac{\ddot a_1}{a_1} - \frac{\ddot a_2}{a_2} -
\frac{\dot a_1}{a_1}\frac{\dot a_2}{a_2} - \frac{m^2}{a_3^2}, \label{ET33}\\
G_0^0 &=&  -\frac{\dot a_1}{a_1}\frac{\dot a_2}{a_2} - \frac{\dot
a_2}{a_2}\frac{\dot a_3}{a_3} - \frac{\dot a_3}{a_3}\frac{\dot
a_1}{a_1} + \frac{m^2}{a_3^2}, \label{ET00}\\
G_3^0 &=&  -  m \left(\frac{\dot a_1}{a_1} -
 \frac{\dot a_2}{a_2}\right). \label{ET03}
\end{eqnarray}
\end{subequations}

\subsection{Spinor field}

For a spinor field $\p$, the symmetry between $\p$ and $\bp$ appears
to demand that one should choose the symmetrized Lagrangian
\cite{kibble}. Keeping this in mind we choose the spinor field
Lagrangian as \cite{saha2001a}:

\begin{equation}
L_{\rm sp} = \frac{\imath}{2} \biggl[\bp \gamma^{\mu} \nabla_{\mu}
\psi- \nabla_{\mu} \bar \psi \gamma^{\mu} \psi \biggr] - m_{\rm sp}
\bp \psi - F, \label{lspin}
\end{equation}
where the nonlinear term $F$ describes the self-interaction of a
spinor field and can be presented as some arbitrary functions of
invariants generated from the real bilinear forms of a spinor field.
Since $\psi$ and $\psi^{\star}$ (complex conjugate of $\psi$) have
four component each, one can construct $4\times 4 = 16$ independent
bilinear combinations. They are
\begin{subequations}
\label{bf}
\begin{eqnarray}
 S&=& \bar \psi \psi\qquad ({\rm scalar}),   \\
  P&=& \imath \bar \psi \gamma^5 \psi\qquad ({\rm pseudoscalar}), \\
 v^\mu &=& (\bar \psi \gamma^\mu \psi) \qquad ({\rm vector}),\\
 A^\mu &=&(\bar \psi \gamma^5 \gamma^\mu \psi)\qquad ({\rm pseudovector}), \\
Q^{\mu\nu} &=&(\bar \psi \sigma^{\mu\nu} \psi)\qquad ({\rm
antisymmetric\,\,\, tensor}),
\end{eqnarray}
\end{subequations}
where $\sigma^{\mu\nu}\,=\,(\imath/2)[\gamma^\mu\gamma^\nu\,-\,
\gamma^\nu\gamma^\mu]$. Invariants, corresponding to the bilinear
forms, are
\begin{subequations}
\label{invariants}
\begin{eqnarray}
I &=& S^2, \\
J &=& P^2, \\
I_v &=& v_\mu\,v^\mu\,=\,(\bar \psi \gamma^\mu \psi)\,g_{\mu\nu}
(\bar \psi \gamma^\nu \psi),\\
I_A &=& A_\mu\,A^\mu\,=\,(\bar \psi \gamma^5 \gamma^\mu
\psi)\,g_{\mu\nu}
(\bar \psi \gamma^5 \gamma^\nu \psi), \\
I_Q &=& Q_{\mu\nu}\,Q^{\mu\nu}\,=\,(\bar \psi \sigma^{\mu\nu}
\psi)\, g_{\mu\alpha}g_{\nu\beta}(\bar \psi \sigma^{\alpha\beta}
\psi).
\end{eqnarray}
\end{subequations}

According to the Fierz identity,  among the five invariants only $I$
and $J$ are independent as all others can be expressed by them: $I_v
= - I_A = I + J$ and $I_Q = I - J.$ Therefore, we choose the
nonlinear term $F$ to be the function of $I$ and $J$ only, i.e., $F
= F(I, J)$, thus claiming that it describes the nonlinearity in its
most general form. Indeed, without losing generality we can choose
$F = F(K)$, with $K$ taking any of the following expressions
$\{I,\,J,\,I+J,\,I-J\}$. Here $\nabla_\mu$ is the covariant
derivative of spinor field:
\begin{equation}
\nabla_\mu \psi = \frac{\partial \psi}{\partial x^\mu} -\G_\mu \psi,
\quad \nabla_\mu \bp = \frac{\partial \bp}{\partial x^\mu} + \bp
\G_\mu, \label{covder}
\end{equation}
with $\G_\mu$ being the spinor affine connection. In \eqref{lspin}
$\gamma$'s are the Dirac matrices in curve space-time and obey the
following algebra
\begin{equation}
\gamma^\mu \gamma^\nu + \gamma^\nu \gamma^\mu = 2 g^{\mu\nu}
\label{al}
\end{equation}
and are connected with the flat space-time Dirac matrices $\bg$ in
the following way
\begin{equation}
 g_{\mu \nu} (x)= e_{\mu}^{a}(x) e_{\nu}^{b}(x) \eta_{ab},
\quad \gamma_\mu(x)= e_{\mu}^{a}(x) \bg_a \label{dg}
\end{equation}
where $e_{\mu}^{a}$ is a set of tetrad 4-vectors.

For the metric \eqref{bvi0} we choose the tetrad as follows:

\begin{equation}
e_0^{(0)} = 1, \quad e_1^{(1)} = a_1 e^{-mx_3}, \quad e_2^{(2)} =
a_2 e^{mx_3}, \quad e_3^{(3)} = a_3. \label{tetradvi}
\end{equation}

The Dirac matrices $\gamma^\mu(x)$ of Bianchi type-$VI_0$ spacetime
are connected with those of Minkowski one as follows:
$$ \gamma^0=\bg^0,\quad \gamma^1 = \frac{ e^{m x_3}}{a_1} \bg^1,
\quad \gamma^2= \frac{ e^{-m x_3}}{a_2}\bg^2,\quad \gamma^3 = \frac{
1}{a_3}\bg^3$$

$$\gamma^5 = - \imath \sqrt{-g}
\gamma^0\gamma^1\gamma^2\gamma^3 = - \imath \bg^0\bg^1\bg^2\bg^3 =
\bg^5
$$
with
\begin{eqnarray}
\bg^0\,=\,\left(\begin{array}{cc}I&0\\0&-I\end{array}\right), \quad
\bg^i\,=\,\left(\begin{array}{cc}0&\sigma^i\\
-\sigma^i&0\end{array}\right), \quad
\gamma^5 = \bg^5&=&\left(\begin{array}{cc}0&-I\\
-I&0\end{array}\right),\nonumber
\end{eqnarray}
where $\sigma_i$ are the Pauli matrices:
\begin{eqnarray}
\sigma^1\,=\,\left(\begin{array}{cc}0&1\\1&0\end{array}\right),
\quad \sigma^2\,=\,\left(\begin{array}{cc}0&-\imath\\
\imath&0\end{array}\right), \quad
\sigma^3\,=\,\left(\begin{array}{cc}1&0\\0&-1\end{array}\right).
\nonumber
\end{eqnarray}
Note that the $\bg$ and the $\sigma$ matrices obey the following
properties:
\begin{eqnarray}
\bg^i \bg^j + \bg^j \bg^i = 2 \eta^{ij},\quad i,j = 0,1,2,3
\nonumber\\
\bg^i \bg^5 + \bg^5 \bg^i = 0, \quad (\bg^5)^2 = I,
\quad i=0,1,2,3 \nonumber\\
\sigma^j \sigma^k = \delta_{jk} + i \varepsilon_{jkl} \sigma^l,
\quad j,k,l = 1,2,3 \nonumber
\end{eqnarray}
where $\eta_{ij} = \{1,-1,-1,-1\}$ is the diagonal matrix,
$\delta_{jk}$ is the Kronekar symbol and $\varepsilon_{jkl}$ is the
totally antisymmetric matrix with $\varepsilon_{123} = +1$.

The spinor affine connection matrices $\G_\mu (x)$ are uniquely
determined up to an additive multiple of the unit matrix by the
equation
\begin{equation}
\frac{\pr \gamma_\nu}{\pr x^\mu} - \G_{\nu\mu}^{\rho}\gamma_\rho -
\G_\mu \gamma_\nu + \gamma_\nu \G_\mu = 0, \label{afsp}
\end{equation}
with the solution
\begin{equation}
\Gamma_\mu = \frac{1}{4} \bg_{a} \gamma^\nu \partial_\mu e^{(a)}_\nu
- \frac{1}{4} \gamma_\rho \gamma^\nu \Gamma^{\rho}_{\mu\nu}.
\label{sfc}
\end{equation}

From the Bianchi type-VI metric \eqref{sfc} one finds the following
expressions for spinor affine connections:
\begin{subequations}
\label{sac123}
\begin{eqnarray}
\G_0 &=& 0, \label{sac0}\\  \G_1 &=& \frac{1}{2}\Bigl(\dot a_1
\bg^1\bg^0 - m\frac{a_1}{a_3} \bg^1\bg^3\Bigr) e^{-mx_3},
\label{sac1}\\  \G_2 &=& \frac{1}{2}\Bigl(\dot a_2 \bg^2\bg^0 +
m\frac{a_2}{a_3} \bg^2\bg^3\Bigr) e^{mx_3}, \label{sac2}\\  \G_3 &=&
\frac{\dot a_3}{2} \bg^3 \bg^0. \label{sac3}
\end{eqnarray}
\end{subequations}

\subsection{Field equations}

Variation of \eqref{action} with respect to the metric function
$g_{\mu \nu}$ gives the Einstein field equation
\begin{equation}
G_\mu^\nu = R_\mu^\nu - \frac{1}{2} \delta_\mu^\nu R = -\kappa
T_\mu^\nu, \label{EEg}
\end{equation}
where $R_\mu^\nu$ and $R$ are the Ricci tensor and Ricci scalar,
respectively. Here $T_\mu^\nu$ is the energy momentum tensor of the
spinor field.

Varying \eqref{lspin} with respect to $\bp (\psi)$ one finds the
spinor field equations:
\begin{subequations}
\label{speq}
\begin{eqnarray}
\imath\gamma^\mu \nabla_\mu \psi - m_{\rm sp} \psi - {\cD} \psi -
 \imath {\cG} \gamma^5 \psi &=&0, \label{speq1} \\
\imath \nabla_\mu \bp \gamma^\mu +  m_{\rm sp} \bp + {\cD}\bp +
\imath {\cG} \bp \gamma^5 &=& 0, \label{speq2}
\end{eqnarray}
\end{subequations}
where we denote ${\cD} = 2 S F_K K_I$ and ${\cG} = 2 P F_K K_J$,
with $F_K = dF/dK$, $K_I = dK/dI$ and $K_J = dK/dJ.$ In view of
\eqref{speq}, eq. \eqref{lspin} can be rewritten as
\begin{eqnarray}
L_{\rm sp} & = & \frac{\imath}{2} \bigl[\bp \gamma^{\mu}
\nabla_{\mu} \psi- \nabla_{\mu} \bar \psi \gamma^{\mu} \psi \bigr] -
m_{\rm sp} \bp \psi - F(I,\,J)
\nonumber \\
& = & \frac{\imath}{2} \bp [\gamma^{\mu} \nabla_{\mu} \psi - m_{\rm
sp} \psi] - \frac{\imath}{2}[\nabla_{\mu} \bar \psi \gamma^{\mu} +
m_{\rm sp} \bp] \psi
- F(I,\,J),\nonumber \\
& = & 2 (I F_I + J F_J) - F = 2 K F_K - F(K). \label{lspin01}
\end{eqnarray}
In what follows we consider the case when the spinor field depends
on $t$ only, i.e.  $\psi = \psi (t)$.

\subsection{Energy momentum tensor of the spinor field}

The energy-momentum tensor of the spinor field is given by
\begin{equation}
T_{\mu}^{\rho}=\frac{\imath}{4} g^{\rho\nu} \biggl(\bp \gamma_\mu
\nabla_\nu \psi + \bp \gamma_\nu \nabla_\mu \psi - \nabla_\mu \bar
\psi \gamma_\nu \psi - \nabla_\nu \bp \gamma_\mu \psi \biggr) \,-
\delta_{\mu}^{\rho} L_{\rm sp}. \label{temsp}
\end{equation}

Then in view of \eqref{covder} and \eqref{lspin01} the
energy-momentum tensor of the spinor field can be written as
\begin{eqnarray}
T_{\mu}^{\,\,\,\rho}&=&\frac{\imath}{4} g^{\rho\nu} \bigl(\bp
\gamma_\mu
\partial_\nu \psi + \bp \gamma_\nu \partial_\mu \psi -
\partial_\mu \bar \psi \gamma_\nu \psi - \partial_\nu \bp \gamma_\mu
\psi \bigr)\nonumber\\
& - &\frac{\imath}{4} g^{\rho\nu} \bp \bigl(\gamma_\mu \G_\nu +
\G_\nu \gamma_\mu + \gamma_\nu \G_\mu + \G_\mu \gamma_\nu\bigr)\psi
 \,- \delta_{\mu}^{\rho} \bigl(2 K F_K - F(K)\bigr). \label{temsp0}
\end{eqnarray}
As is seen from \eqref{temsp0}, in the case where, for a given
metric $\G_\mu$'s are different, there arise nontrivial non-diagonal
components of the energy momentum tensor.

After a little manipulations from \eqref{temsp0} one finds the
following components of the energy momentum tensor:
\begin{subequations}
\label{Ttot}
\begin{eqnarray}
T_0^0 & = & m_{\rm sp} S + F(K), \label{emt00}\\
T_1^1 &=& T_2^2 = T_3^3 =  F(K) - 2 K F_K, \label{emtii}\\
T_1^0 &=& -\frac{\imath}{4} m \frac{a_1}{a_3} e^{-m x_3}\, \bp \bg^3
\bg^1 \bg^0 \psi
= -\frac{1}{4} m \frac{a_1}{a_3} e^{-m x_3}\, A^2 , \label{emt01} \\
T_2^0 &=&-\frac{\imath}{4} m \frac{a_2}{a_3} e^{m x_3}\, \bp \bg^2
\bg^3 \bg^0 \psi
= -\frac{1}{4} m \frac{a_2}{a_3} e^{m x_3}\,A^1, \label{emt02} \\
T_3^0 &=& 0, \label{emt03} \\
T_2^1 &=& \frac{\imath}{4} \frac{a_2}{a_1} e^{2m x_3}
\biggl[\biggl(\frac{\dot a_1}{a_1} - \frac{\dot a_2}{a_2}\biggr) \bp
\bg^1 \bg^2 \bg^0 \psi - \frac{2m}{a_3} \bp \bg^1 \bg^2 \bg^3 \psi
\biggr] \nonumber \\
&=& \frac{1}{4} \frac{a_2}{a_1} e^{2m x_3} \biggl[\biggl(\frac{\dot
a_1}{a_1} - \frac{\dot a_2}{a_2}\biggr) A^3
- \frac{2m}{a_3}A^0\biggr] , \label{emt12}\\
T_3^1 &=&\frac{\imath}{4} \frac{a_3}{a_1} e^{m x_3}
\biggl(\frac{\dot a_3}{a_3} - \frac{\dot a_1}{a_1}\biggr) \bp \bg^3
\bg^1 \bg^0 \psi = \frac{1}{4} \frac{a_3}{a_1} e^{m x_3}
\biggl(\frac{\dot a_3}{a_3} - \frac{\dot a_1}{a_1}\biggr) A^2 \label{emt13}\\
T_3^2 &=&\frac{\imath}{4} \frac{a_3}{a_2} e^{-m x_3}
\biggl(\frac{\dot a_2}{a_2} - \frac{\dot a_3}{a_3}\biggr) \bp \bg^2
\bg^3 \bg^0 \psi = \frac{1}{4} \frac{a_3}{a_2} e^{-m x_3}
\biggl(\frac{\dot a_2}{a_2} - \frac{\dot a_3}{a_3}\biggr)A^1.
\label{emt23}
\end{eqnarray}
\end{subequations}

As one sees from \eqref{Ttot} the spinor field possesses non-trivial
of non-diagonal components of the energy momentum tensor.

\section{Solution to the field equations}

In this section we solve the self consistent system of spinor and
gravitational field equations. We begin with the spinor field
equations and then solve the gravitational field equations. Finally
we study the influence of the non-diagonal components of the energy
momentum tensor on the components of the spinor field and metric
functions.

\subsection{Solution to the spinor field equation}

Let us begin with the spinor field equations. In view of
\eqref{covder} and \eqref{sac123} the spinor field equations
\eqref{speq} take the form

\begin{subequations}
\label{SF1}
\begin{eqnarray}
\imath \bg^0 \bigl(\dot \psi + \frac{1}{2}\frac{\dot V}{V}
\psi\bigr) - m_{\rm sp} \psi - {\cD}
\psi -  \imath {\cG} \bg^5 \psi &=&0, \label{speq1p}\\
\imath \bigl(\dot \bp + \frac{1}{2}\frac{\dot V}{V} \bp\bigr)\bg^0 +
m_{\rm sp} \bp  + {\cD}  \bp + \imath {\cG}\bp \bg^5 &=& 0,
\label{speq2p}
\end{eqnarray}
\end{subequations}
where we define the volume scale
\begin{equation}
V = a_1 a_2 a_3. \label{VDef}
\end{equation}

As we have already mentioned, $\psi$ is a function of $t$ only. We
consider the 4-component spinor field given by
\begin{eqnarray}
\psi = \left(\begin{array}{c} \psi_1\\ \psi_2\\ \psi_3 \\
\psi_4\end{array}\right). \label{psi}
\end{eqnarray}
Denoting $\phi_i =\sqrt{V} \psi_i$  from \eqref{speq1p} for the
spinor field we find we find
\begin{subequations}
\label{speq1pfg}
\begin{eqnarray}
\dot \phi_1 + \imath\, {\Phi} \phi_1 + {\cG} \phi_3 &=& 0, \label{ph1}\\
\dot \phi_2 + \imath\, {\Phi} \phi_2 + {\cG} \phi_4 &=& 0, \label{ph2}\\
\dot \phi_3 - \imath\, {\Phi} \phi_3 -  {\cG}\phi_1 &=& 0, \label{ph3}\\
\dot \phi_4 - \imath\, {\Phi} \phi_4 -  {\cG}\phi_2&=& 0.
\label{ph4}
\end{eqnarray}
\end{subequations}
The foregoing system of equations can be written in the form:
\begin{equation}
\dot \phi = A \phi, \label{phi}
\end{equation}
with $\phi = {\rm
col}\left(\phi_1,\,\phi_2,\,\phi_3,\,\phi_4\right)$ and
\begin{equation}
A = \left(\begin{array}{cccc}-\imath\, \Phi &0 & - {\cG}& 0 \\
0&-\imath\, \Phi& 0 & - {\cG}\\ {\cG}&0&\imath\, \Phi&0\\0& {\cG} &
0 &\imath\, \Phi
\end{array}\right). \label{AMat}
\end{equation}
It can be easily found that
\begin{equation}
{\rm det} A = \left(\Phi^2  +{\cG}^2\right)^2. \label{detA}
\end{equation}

The solution to the equation \eqref{phi} can be written in the form
\begin{equation}
\phi(t) = {\rm T exp}\Bigl(-\int_t^{t_1}  A_1 (\tau) d \tau\Bigr)
\phi (t_1), \label{phi1}
\end{equation}
where
\begin{equation}
A_1 = \left(\begin{array}{cccc}-\imath\, {\cD} &0 & - {\cG}& 0 \\
0&-\imath\, {\cD}& 0 & - {\cG}\\
{\cG}&0&\imath\, {\cD}&0\\0&{\cG} & 0 &\imath\, {\cD}
\end{array}\right). \label{AMat1}
\end{equation}
and $\phi (t_1)$ is the solution at $t = t_1$, with $t_1$ being
quite large, so that the volume scale $V$, hence the expanding
Universe becomes large enough. As it will be shown later, $K =
V_0^2/V^2$ for $K$ taking one of the following expressions
$\{J,\,I+J,\,I-J\}$ with trivial spinor-mass and $K = V_0^2/V^2$ for
$K = I$ for any spinor-mass. Since our Universe is expanding, the
quantities ${\cD}$ and ${\cG}$ become trivial at large $t = t_1$.
Hence in case of  $K = I$ with non-trivial spinor-mass one can
assume $\phi (t_1) = {\rm col}\left(e^{-\imath m_{\rm sp}
t_1},\,e^{-\imath m_{\rm sp} t_1},\,e^{\imath m_{\rm sp}
t_1},\,e^{\imath m_{\rm sp} t_1}\right)$, whereas for other cases
with trivial spinor-mass we have $\phi (t_1) = {\rm
col}\left(\phi_{1}^{0},\,\phi_{2}^{0},\,\phi_{3}^{0},\,\phi_{4}^{0}\right)$
with $\phi_i^0$ being some constants. Here we have used the fact
that $\Phi = m_{\rm sp} + {\cD}.$ The other way to solve the system
\eqref{speq1pfg} is given in \cite{saha2004b}.

It can be shown that bilinear spinor forms \eqref{bf} the obey the
following system of equations:
\begin{subequations}
\label{inv}
\begin{eqnarray}
\dot S_0  +  {\cG} A_{0}^{0} &=& 0, \label{S0} \\
\dot P_0  -  \Phi A_{0}^{0} &=& 0, \label{P0}\\
\dot A_{0}^{0} +  \Phi P_0 -  {\cG}
S_0 &=& 0, \label{A00}\\
\dot A_{0}^{3}  &=& 0, \label{A03}\\
\dot v_{0}^{0}  &=& 0,\label{v00} \\
\dot v_{0}^{3} +
\Phi Q_{0}^{30} +  {\cG} Q_{0}^{21} &=& 0,\label{v03}\\
\dot Q_{0}^{30}  -  \Phi v_{0}^{3} &=& 0,\label{Q030} \\
\dot Q_{0}^{21}  -  {\cG} v_{0}^{3} &=& 0, \label{Q021}
\end{eqnarray}
\end{subequations}
where we denote $S_0 = S V$, $P_0 = P V$, $A_0^\mu = A^\mu V$,
$v_0^\mu = v^\mu V$ and $Q_0^{\mu \nu} = Q^{\mu \nu} V$ . Combining
these equations together and taking the first integral one gets
\begin{subequations}
\label{inv0}
\begin{eqnarray}
(S_{0})^{2} + (P_{0})^{2} + (A_{0}^{0})^{2}  &=&
l_1^2 = {\rm Const}, \label{inv01}\\
A_{0}^{3} &=& l_2^2 = {\rm Const}, \label{inv02}\\
 (Q_{0}^{30})^{2} + (Q_{0}^{21})^{2} +
(v_{0}^{3})^{2} &=& l_3^2 = {\rm Const}, \label{inv03} \\
v_{0}^{0} &=& l_4^2 = {\rm Const}. \label{inv04}
\end{eqnarray}
\end{subequations}

\subsection{Solution to the gravitational field equation}

Now let us consider the gravitational field equations. In view of
\eqref{ET} and \eqref{Ttot} with find the following system of
Einstein Equations

\begin{subequations}
\label{EEbvi0}
\begin{eqnarray}
\frac{\ddot a_2}{a_2} + \frac{\ddot a_3}{a_3} +
\frac{\dot a_2}{a_2}\frac{\dot a_3}{a_3} - \frac{m^2}{a_3^2} &=& \kappa\bigl(F(K) - 2 K F_K\bigr), \label{EE11}\\
\frac{\ddot a_3}{a_3} + \frac{\ddot a_1}{a_1} +
\frac{\dot a_3}{a_3}\frac{\dot a_1}{a_1} - \frac{m^2}{a_3^2} &=& \kappa\bigl(F(K) - 2 K F_K\bigr), \label{EE22}\\
\frac{\ddot a_1}{a_1} + \frac{\ddot a_2}{a_2} +
\frac{\dot a_1}{a_1}\frac{\dot a_2}{a_2} + \frac{m^2}{a_3^2} &=& \kappa\bigl(F(K) - 2 K F_K\bigr), \label{EE33}\\
\frac{\dot a_1}{a_1}\frac{\dot a_2}{a_2} + \frac{\dot
a_2}{a_2}\frac{\dot a_3}{a_3} + \frac{\dot a_3}{a_3}\frac{\dot
a_1}{a_1} - \frac{m^2}{a_3^2} &=&   \kappa\bigl(m_{\rm sp} S + F(K)\bigr), \label{EE00}\\
\frac{\dot a_1}{a_1} -  \frac{\dot a_2}{a_2}  &=& 0, \label{EE03}
\end{eqnarray}
\end{subequations}
with the additional constraints
\begin{subequations}
\label{AC}
\begin{eqnarray}
T_1^0 &=& -\frac{1}{4} m \frac{a_1}{a_3} e^{-m x_3}\, A^2 = 0, \label{AC01} \\
T_2^0 &=&-\frac{1}{4} m \frac{a_2}{a_3} e^{m x_3}\,A^1 = 0, \label{AC02} \\
T_2^1 &=& \frac{1}{4} \frac{a_2}{a_1} e^{2m x_3}
\biggl[\biggl(\frac{\dot a_1}{a_1} - \frac{\dot a_2}{a_2}\biggr) A^3
- \frac{2m}{a_3}A^0\biggr] = 0, \label{AC12}\\
T_2^1 &=& \frac{1}{4} \frac{a_3}{a_1} e^{m x_3}
\biggl(\frac{\dot a_3}{a_3} - \frac{\dot a_1}{a_1}\biggr) A^2 = 0, \label{AC13}\\
T_3^2 &=&\frac{1}{4} \frac{a_3}{a_2} e^{-m x_3} \biggl(\frac{\dot
a_2}{a_2} - \frac{\dot a_3}{a_3}\biggr)A^1 = 0. \label{AC23}
\end{eqnarray}
\end{subequations}

From \eqref{EE03} one dully finds

\begin{equation}
a_2 = X_0 a_1, \quad X_0 = {\rm Const}. \label{a21}
\end{equation}

Let us now find expansion and shear for Bianchi type-$VI_0$ metric.
The expansion is given by
\begin{equation}
\vartheta = u^\mu_{;\mu} = u^\mu_{\mu} + \G^\mu_{\mu\alpha}
u^\alpha, \label{expansion}
\end{equation}
and the shear is given by
\begin{equation}
\sigma^2 = \frac{1}{2} \sigma_{\mu\nu} \sigma^{\mu\nu},
\label{shear}
\end{equation}
with
\begin{equation}
\sigma_{\mu\nu} = \frac{1}{2}\bigl[u_{\mu;\alpha} P^\alpha_\nu +
u_{\nu;\alpha} P^\alpha_\mu \bigr] - \frac{1}{3} \vartheta
P_{\mu\nu}, \label{shearcomp}
\end{equation}
where the projection vector $P$:
\begin{equation}
P^2 = P, \quad P_{\mu\nu} = g_{\mu\nu} - u_\mu u_\nu, \quad
P^\mu_\nu = \delta^\mu_\nu - u^\mu u_\nu. \label{proj}
\end{equation}
In comoving system we have $u^\mu = (1,0,0,0)$. In this case one
finds
\begin{equation}
\vartheta = \frac{\dot a_1}{a_1} + \frac{\dot a_2}{a_2} + \frac{\dot
a_3}{a_3} = \frac{\dot V}{V}, \label{expbvi0}
\end{equation}
and
\begin{subequations}
\label{shearcomps}
\begin{eqnarray}
\sigma_{1}^{1} &=& -\frac{1}{3}\Bigl(-2\frac{\dot a_1}{a_1} +
\frac{\dot
a_2}{a_2} + \frac{\dot a_3}{a_3}\Bigr) =  \frac{\dot a_1}{a_1} - \frac{1}{3} \vartheta, \label{sh11}\\
\sigma_{2}^{2} &=& -\frac{1}{3}\Bigl(-2\frac{\dot a_2}{a_2} +
\frac{\dot a_3}{a_3} +
\frac{\dot a_1}{a_1}\Bigr) =  \frac{\dot a_2}{a_2} - \frac{1}{3} \vartheta, \label{sh22}\\
\sigma_{3}^{3} &=& -\frac{1}{3}\Bigl(-2\frac{\dot a_3}{a_3} +
\frac{\dot a_1}{a_1} + \frac{\dot a_2}{a_2}\Bigr) =  \frac{\dot
a_3}{a_3} - \frac{1}{3} \vartheta. \label{sh33}
\end{eqnarray}
\end{subequations}

One then finds
\begin{equation}
\sigma^ 2 = \frac{1}{2}\biggl[\sum_{i=1}^3 \biggl(\frac{\dot
a_i}{a_i}\biggr)^2 - \frac{1}{3}\vartheta^2\biggr] =
\frac{1}{2}\biggl[\sum_{i=1}^3 H_i^2 -
\frac{1}{3}\vartheta^2\biggr]. \label{sheargen}
\end{equation}

Inserting \eqref{a21}  into \eqref{expbvi0} and \eqref{shearcomps}
we find

\begin{equation}
\vartheta =2 \frac{\dot a_1}{a_1} + \frac{\dot a_3}{a_3},
\label{expbvi1}
\end{equation}
and
\begin{subequations}
\label{shearcomps0}
\begin{eqnarray}
\sigma_{1}^{1} &=& \frac{1}{3}\Bigl(\frac{\dot a_1}{a_1} -
\frac{\dot a_3}{a_3}\Bigr), \label{sh110}\\
\sigma_{2}^{2} &=&  \frac{1}{3}\Bigl(\frac{\dot a_1}{a_1}
- \frac{\dot a_3}{a_3}\Bigr), \label{sh220}\\
\sigma_{3}^{3} &=& - \frac{2}{3}\Bigl(\frac{\dot a_1}{a_1} -
\frac{\dot a_3}{a_3}\Bigr). \label{sh33a}
\end{eqnarray}
\end{subequations}
As it was found in previous papers, due to explicit presence of
$a_3$ in the Einstein equations, one needs some additional
conditions. In an early work we propose two different situation,
namely, set $a_3 = \sqrt{V}$ and $a_3 = V$ which allowed us to
obtain exact solutions for the metric functions.

In a recent paper we imposed the proportionality condition, widely
used in literature. Demanding that the expansion is proportion to a
component of the shear tensor, namely
\begin{equation}
\vartheta = N_3 \sigma_3^3.\label{propconvi}
\end{equation}
The motivation behind assuming this condition is explained with
reference to  Thorne \cite{thorne67}. The observations of the
velocity-red-shift relation for extragalactic sources suggest that
Hubble expansion of the universe is isotropic today within $\approx
30$ per cent \cite{kans66,ks66}. To put more precisely, red-shift
studies place the limit
\begin{equation}
\frac{\sigma}{H} \leq 0.3, \label{propconviexp}
\end{equation}
on the ratio of shear $\sigma$ to Hubble constant $H$ in the
neighborhood of our Galaxy today. Collins et al. \cite{Collins} have
pointed out that for spatially homogeneous metric, the normal
congruence to the homogeneous hypersurfaces satisfies the condition:
$\frac{\sigma}{\theta} = {\rm const.}$ Under this proportionality
condition it was also found that the energy-momentum distribution of
the model is strictly isotropic, which is absolutely true for our
case.

Further on account of \eqref{VDef} we finally find
\begin{eqnarray}
a_1 = \Bigl[\frac{1}{X_0 X_1} V\Bigr]^{\frac{1}{3}- \frac{1}{2N_3}},
\quad a_2 = X_0 \Bigl[\frac{1}{X_0 X_1} V\Bigr]^{\frac{1}{3}-
\frac{1}{2N_3}},\quad a_3 = X_1 \Bigl[\frac{1}{X_0 X_1}
V\Bigr]^{\frac{1}{3}+ \frac{1}{N_3}}. \label{Metf}
\end{eqnarray}
As it is obvious from \eqref{Metf} the isotropization of the
spacetime can take place only for large value of $N_3$.

The equation for $V$ can be found from the Einstein Equation
\eqref{ET} which after some manipulation looks
\begin{equation}
\ddot V = \bar X V^{1/3 - 2/N_3} + \frac{3 \kappa}{2} \bigl[m_{\rm
sp} S + 2 \bigl(F(K) - K F_K\bigr)\bigr] V,  \label{Vdefein}
\end{equation}
with $\bar X = 2 m^2 X_0^{2/N_3 + 2/3} X_1^{2/N_3 - 1/3}.$ In order
to solve \eqref{Vdefein} we have to know the relation between $K$
and $V$. Recalling that  $K$ takes one of the following expressions
$\{I,\,J,\,I+J,\,I-J\}$, with ${\cD} = 2 S F_K K_I$ and ${\cG} = 2 P
F_K K_J$ let us first find those relations for different $K$. \vskip
5mm

In case of $K = I$, i.e. ${\cG} = 0$ from \eqref{S0} we find
\begin{equation}
\dot S_0 = 0, \label{S0n}
\end{equation}
with the solution
\begin{equation}
K = I = S^2 = \frac{V_0^2}{V^2}, \quad \Rightarrow \quad S =
\frac{V_0}{V}, \quad V_0 = {\rm const.} \label{SV}
\end{equation}
In  this case spinor field can be either massive or massless.

In the cases where $K$ takes any of the following expressions
$\{J,\,I+J,\,I-J\}$ that gives $K_J = \pm 1$, we consider a massless
spinor field. \vskip 5mm

In case of $K = J$, \quad $ \Phi = {\cD} = 0$. Then from \eqref{P0}
we have
\begin{equation}
\dot P_0 = 0, \label{P0n}
\end{equation}
with the solution
\begin{equation}
K = J = P^2 = \frac{V_0^2}{V^2}, \quad \Rightarrow \quad P =
\frac{V_0}{V}, \quad V_0 = {\rm const.} \label{PV}
\end{equation}

\vskip 5mm

In case of $K = I + J$ the equations \eqref{S0} and \eqref{P0} can
be rewritten as
\begin{subequations}
\begin{eqnarray}
\dot S_0  +  2 P F_K  A_{0}^{0} &=& 0, \label{S0new} \\
\dot P_0  -  2 S F_K  A_{0}^{0} &=& 0, \label{P0new}
\end{eqnarray}
\end{subequations}
which can be rearranged as
\begin{equation}
S_0 \dot S_0 +  P_0 \dot P_0 = \frac{1}{2}\frac{d}{dt}\bigl( S_0^2 +
P_0^2\bigr) = \frac{1}{2} \frac{d}{dt}\bigl(V^2 K\bigr) = 0,
\label{K0}
\end{equation}
with the solution
\begin{equation}
K = I + J = S^2 + P^2 = \frac{V_0^2}{V^2}, \quad V_0 = {\rm const.}
\label{KV}
\end{equation}

It should be noticed that in this case one can use the following
parametrization for $S$ and $P$:

\begin{equation}
S = \sqrt{K} \sin \theta = \frac{V_0}{V} \sin \theta, \quad   P =
\sqrt{K} \cos \theta = \frac{V_0}{V} \cos \theta. \label{KIpJ}
\end{equation}

Here we like to note that for the case in concern one can consider
the massive spinor field as well. In that case we have
\begin{subequations}
\begin{eqnarray}
\dot S_0  +  2 P F_K  A_{0}^{0} &=& 0, \label{S0new1} \\
\dot P_0 - m_{\rm sp} A_{0}^{0} -  2 S F_K  A_{0}^{0} &=& 0,
\label{P0new1}
\end{eqnarray}
\end{subequations}
which can be rearranged as
\begin{equation}
S_0 \dot S_0 +  P_0 \dot P_0 = \frac{1}{2} \frac{d}{dt}\bigl(S_0^2 +
P_0^2\bigr) = \frac{1}{2}\frac{d}{dt}\bigl(V^2 K\bigr) =  m_{\rm sp}
P_0 A_{0}^{0}. \label{K01}
\end{equation}

From \eqref{inv01} follows
\begin{equation}
(A_0^0)^2 = l_1^2 - V^2\left(S^2 + P^2\right) = l_1^2 - V^2 K.
\label{KVA0}
\end{equation}
Further setting $S = \sqrt{K} \sin \theta$ and $P = \sqrt{K} \cos
\theta$ Eq. \eqref{K01} can be written as
\begin{equation}
\frac{d\left(V^2 K\right)}{\sqrt{\left(V^2 K\right)\left( l_1^2 -
\left(V^2 K\right)\right)}} = 2  m_{\rm sp} \cos \theta dt,
\label{KVA1}
\end{equation}
with the solution
\begin{equation}
K =  \frac{l_1^2}{2 V^2}\left(1 + \sin\left(2 m_{\rm sp} \cos \theta
t\right) \right). \label{KV1n}
\end{equation}
As one sees for massless spinor field from \eqref{KV1n} follows $K =
l_1^2/2 V^2$, which is equivalent to \eqref{KV} for $V_0^2 =
l_1^2/2$. Moreover, given the fact that $\sin\left(2 m_{\rm sp} \cos
\theta t\right) \in [-1,\,1]$ for the massive spinor field $K$ comes
out to be a time varying quantity that has the range $K \in
[0,\,l_1^2/V^2].$ In our purpose we consider here only the massless
spinor field. \vskip 5mm

Finally, for $K = I - J$ the equations \eqref{S0} and \eqref{P0} can
be rewritten as
\begin{subequations}
\begin{eqnarray}
\dot S_0  -  2 P F_K  A_{0}^{0} &=& 0, \label{S0new1a} \\
\dot P_0  -  2 S F_K  A_{0}^{0} &=& 0, \label{P0new1a}
\end{eqnarray}
\end{subequations}
which can be rearranged as
\begin{equation}
S_0 \dot S_0 -  P_0 \dot P_0 = \frac{d}{dt}\bigl( S_0^2 -
P_0^2\bigr) = \frac{d}{dt}\bigl(V^2 K\bigr) = 0, \label{K02}
\end{equation}
with the solution
\begin{equation}
K = I - J = S^2 - P^2 = \frac{V_0^2}{V^2}, \quad V_0 = {\rm const.}
\label{KV1}
\end{equation}

In this case one can use the following parametrization for $S$ and
$P$:

\begin{equation}
S = \sqrt{K} \cosh \theta = \frac{V_0}{V} \cosh \theta, \quad   P =
\sqrt{K} \sinh \theta = \frac{V_0}{V} \sinh \theta. \label{KImJ}
\end{equation}

Thus we see that $K$ is a function of $V$. For the cases considered
here we established that $K = V_0^2/V^2$. So one can easily consider
the case when $K = I = S^2$. In that case it is possible to study
both massive and massless spinor field to clarify the role of spinor
mass. Further inserting $F(K)$ into \eqref{Vdefein} one finds the
expression for $V$. One can further study the behavior of $V$
numerically for different $F$. But before that let us first see what
happens to the result obtained if the additional conditions are
taken into account. \vskip 5mm

\section{Influence of Additional conditions on the solutions}

Thus, until this point we have only used the Einstein system of
equations without the additional conditions \eqref{AC}. In what
follows we turn to them and see how these conditions effect our
solutions.

From \eqref{AC01} and \eqref{AC02} one dully finds
\begin{equation}
A^2 = 0, \quad {\rm and} \quad A^1 = 0. \label{A12}
\end{equation}
In view of \eqref{A12} the relations \eqref{AC13} and \eqref{AC23}
fulfill even without imposing restrictions on the metric functions.
On account of \eqref{EE03} from \eqref{AC12} one finds
\begin{equation}
A^0 = 0. \label{A00a}
\end{equation}
The equalities \eqref{A12} and \eqref{A00a} can be rewritten in
terms of spinor field components as follows:

\begin{subequations}
\begin{eqnarray}
\psi_1^* \psi_2 - \psi_2^* \psi_1 + \psi_3^* \psi_4 - \psi_4^*
\psi_3 &=& 0, \label{A2n} \\
\psi_1^* \psi_2 + \psi_2^* \psi_1 + \psi_3^* \psi_4 + \psi_4^*
\psi_3 &=& 0, \label{A1n} \\
\psi_1^* \psi_3 + \psi_2^* \psi_4 + \psi_3^* \psi_1 + \psi_4^*
\psi_2 &=& 0. \label{A0n}
\end{eqnarray}
\end{subequations}

On the other hand, in view of \eqref{A12} and \eqref{A00a} from the
equality
\begin{equation}
v_\mu A^\mu = 0 \Longrightarrow v_3 A^3 = 0, \label{A3V3}
\end{equation}
we have either
\begin{equation}
A^3 = 0 \Longrightarrow \psi_1^* \psi_1 - \psi_2^* \psi_2 + \psi_3^*
\psi_3 - \psi_4^* \psi_4 = 0, \label{A3n}
\end{equation}
or
\begin{equation}
v^3 = 0 \Longrightarrow \psi_1^* \psi_3 - \psi_2^* \psi_4 + \psi_3^*
\psi_1 - \psi_4^* \psi_2 = 0. \label{v3n}
\end{equation}
In case of $A^3 = 0$ we find $A^\mu = 0$. Then taking into account
that $I_A = A_\mu A^\mu = - (S^2 + P^2)$ we ultimately find
\cite{sahaAPSS2015}
\begin{equation}
S^2 + P^2 = 0 \Longrightarrow S = 0 \quad {\rm and} \quad P = 0.
\label{SPzero}
\end{equation}
Thus we see that in the case considered here the initially massive,
nonlinear spinor field becomes linear and massless as a result of
special geometry of the Bianchi type-$VI_0$ spacetime, which is
equivalent to solving the corresponding Einstein equation in vacuum.

In this case for volume scale $V$ we find
\begin{equation}
\ddot V = \bar X V^{1/3 - 2/N_3}, \label{Vdefbv0lin}
\end{equation}
with the solution in quadrature
\begin{equation}
\frac{dV}{\Phi_0} = t + t_0, \quad \Phi_0 = \sqrt{\frac{6N_3 \bar
X}{4 N_3 - 6} V^{(4 N_3 - 6)/3N_3} + C_0}, \label{Vbv0linsol}
\end{equation}
with $t_0$ and $C_0$ are being some arbitrary constants.

\begin{figure}[ht]
\centering
\includegraphics[height=70mm]{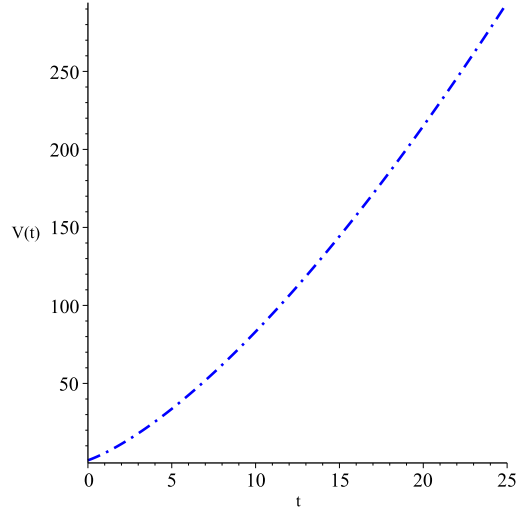} \\
\caption{Evolution of the Universe in absence of spinor field
(Vacuum solution).} \label{Vbv0lin}.
\end{figure}

In Fig. \ref{Vbv0lin} we have plotted the evolution of volume scale.
For simplicity we have set $m = 1$, $X_0 = 1$, $X_1 = 1$, $C_0 = 10$
and $N_3 = 3$. The initial value of volume scale is taken to be
$V(0) = 0.1$ and $\dot V(0)$ is calculated using \eqref{Vbv0linsol}.

As far as spinor field is concerned the Matrix $A$ in \eqref{phi} in
this case becomes trivial and the components of the spinor field can
be written as
\begin{equation}
\psi_i = \frac{c_i}{\sqrt{V}}, \quad i = 1,\,2,\,3,\,4,
\label{psicomp}
\end{equation}
with $c_i$'s being the constant of integration obeying

\begin{subequations}
\begin{eqnarray}
c_1^* c_1 + c_2^* c_2 - c_3^* c_3 -c_4^* c_4 &=& 0, \label{S0new10} \\
c_1^* c_3 + c_2^* c_4 - c_3^* c_1 - c_4^* c_2 &=& 0. \label{P0new10}
\end{eqnarray}
\end{subequations}

The second possibility is to consider \eqref{v3n} with $A^3 \ne 0$.
In this case the nonlinear term as well as the massive term do not
vanish. In what follows, we will consider the case for $K = I$,
setting
\begin{equation}
F = \sum_{k} \lambda_k I^{n_k} =  \sum_{k} \lambda_k S^{2 n_k}.
\label{nonlinearity}
\end{equation}
As far as other cases are concerned we can revive them setting
spinor mass $m_{\rm sp} = 0$.

Then inserting \eqref{nonlinearity} into \eqref{Vdefein} and taking
into account that in this case $S = V_0/V$ we find

\begin{eqnarray}
\ddot V = \Phi_1 (V), \quad \Phi_1 = \bar X V^{1/3 - 2/N_3} +
\frac{3 \kappa}{2} \left[m_{\rm sp}\,V_0 + 2 \sum_{k} \lambda_k( 1 -
n_k) V_0^{2n_k} V^{1 - 2n_k}\right], \label{detvbvi0new}
\end{eqnarray}
with the solution in quadrature
\begin{equation}
\frac{dV}{\Phi_2} = t + t_0, \quad \Phi_2 = \sqrt{\frac{6N_3 \bar
X}{4 N_3 - 6} V^{(4 N_3 - 6)/3N_3} + 3\kappa \left[m_{\rm sp}\,V_0 V
+  \sum_{k} \lambda_k V_0^{2n_k} V^{2(1 - n_k)}\right] + C_1},
\label{Vbv0nonlinsol}
\end{equation}
with $t_0$ and $C_1$ are being some arbitrary constants.

It can be shown that the metric functions and the components of the
spinor field as well as the invariants constructed form them are
inverse function of $V$ of some degree, hence at any spacetime point
where $V = 0$ it is a singularity. So we assume that at the
beginning $V$ was small but non-zero. From \eqref{detvbvi0new} we
see that at initial stage the nonlinear term prevails if $n_k = n_1$
such that $n_1 > 1/2$ and $n_1 > 1/3 + 1/N_3$, whereas for the
nonlinearity to become dominant for large value of $V$ one should
have  $n_k = n_2$ such that $n_2 < 1/2$ and $n_2 < 1/3 + 1/N_3$.

\begin{figure}[ht]
\centering
\includegraphics[height=70mm]{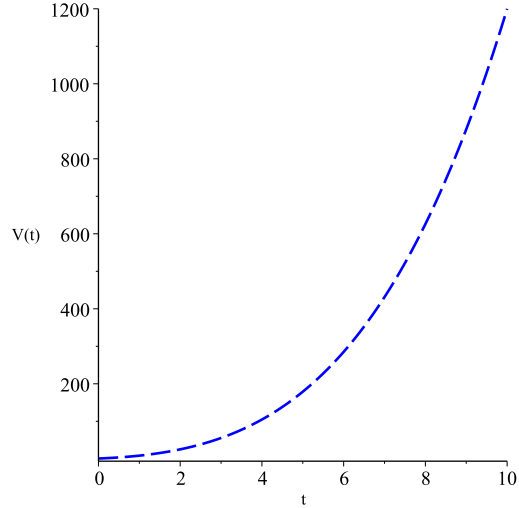} \\
\caption{Evolution of the Universe filled with massive spinor field
 with a positive self-coupling constants  $\lambda_1 = 1$
and $\lambda_2 = 1$.} \label{Vbv0pos}.
\end{figure}

\begin{figure}[ht]
\centering
\includegraphics[height=70mm]{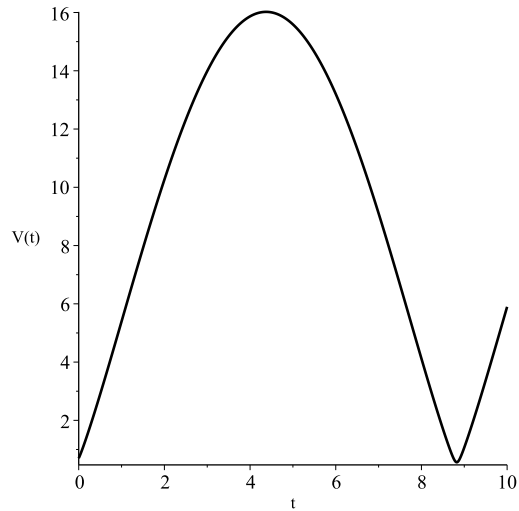} \\
\caption{Evolution of the Universe filled with massive spinor field
with a negative self-coupling constants  $\lambda_1 = -0.5$ and
$\lambda_2 = -0.5$.} \label{Vbv0neg}.
\end{figure}

In Figs. \ref{Vbv0pos} and \ref{Vbv0neg} we have plotted the
evolution of volume scale for positive and negative coupling
constants $\lambda_1$ and $\lambda_2$, respectively. For simplicity
we set $m = 1$, $X_0 = 1$, $X_1 = 1$, $N_3 = 3$, $m_{\rm sp} = 1$,
$V_0 = 1$, $\kappa = 1$, $C_1 = 10$, $n_1 = 3$, $n_2 = 1/4$ and $N_3
= 3$.  In case of positive coupling constants $\lambda_1 = 1$ and
$\lambda_2 = 1$ the model describes an expanding Universe, while for
negative coupling constants $\lambda_1 = -0.5$ and $\lambda_2 = -
0.5$ we have a cyclic Universe that expands to some maximum and then
contracts to minimum only to expand again. The initial value of
volume scale is taken to be $V(0) = 0.1$ and $\dot V(0)$ is
calculated using \eqref{Vbv0nonlinsol}.

Let us also see what happens to deceleration parameter for positive
coupling constants. Using the definition
\begin{equation}
q = - \frac{V \ddot V}{{\dot V}^2} = - \frac{V \Phi_1}{\Phi_2^2}
\label{decpar}
\end{equation}
it can be shown that in this case our Universe is expanding with
acceleration. Taking into account the discussion about the value of
$n_k$ we can rewrite

\begin{equation}
q = - \frac{\bar X V^{4/3 - 2/N_3} + \frac{3 \kappa}{2} \left[m_{\rm
sp}\,V_0 V + 2 \lambda_1( 1 - n_1) V_0^{2n_1} V^{2(1 - n_1)} + 2
\lambda_2( 1 - n_2) V_0^{2n_2} V^{2(1 - n_2)} \right]}{\frac{6N_3
\bar X}{4 N_3 - 6} V^{4/3 - 2/N_3} + 3\kappa \left[m_{\rm sp}\,V_0 V
+  \lambda_1 V_0^{2n_1} V^{2(1 - n_1)} + \lambda_2 V_0^{2n_2} V^{2(1
- n_2)}\right] + C_1}. \label{decparext}
\end{equation}
As it was mentioned earlier, at large $t$, hence for large $V$
prevails the term with  $n_k = n_2 < 1/2$. Taking this into account
we find
\begin{equation}
\lim_{V \to \infty}q  \longrightarrow - (1 - n_2) < 0.
\label{accele}
\end{equation}

\begin{figure}[ht]
\centering
\includegraphics[height=70mm]{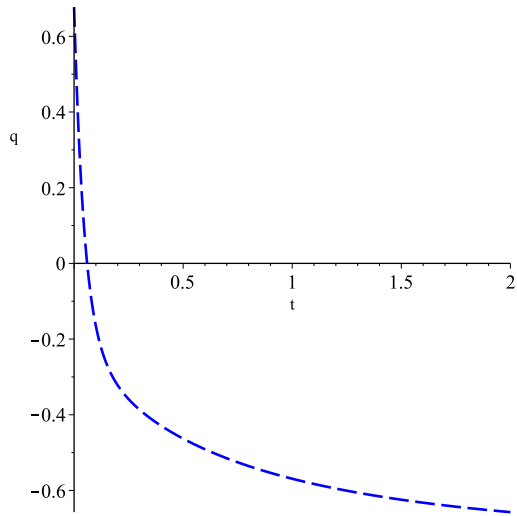} \\
\caption{Evolution of the deceleration parameter for the Universe
filled with massive spinor field with a positive self-coupling
constants  $\lambda_1 = 1$ and $\lambda_2 = 1$.} \label{Vbv0dec}.
\end{figure}

In Fig. \ref{Vbv0dec} we have illustrated the evolution of the the
deceleration parameter. As we see that the spinor field nonlinearity
leads to  the late time accelerated expansion of the Universe.

\section{Conclusion}

Within the scope of Bianchi type-$VI_0$ spacetime we study the role
of spinor field on the evolution of the Universe. In this case we
consider the spinor field that depends only on time $t$. Even in
this case the spinor field possesses non-zero non-diagonal
components of energy-momentum tensor thanks to its specific relation
with gravitational field. This fact plays vital role on the
evolution of the Universe. Due to the specific behavior of the
spinor field we have two different scenarios. In one case the
bilinear forms constructed from it becomes trivial, thus giving rise
to a massless and linear spinor field Lagrangian. This case is
equivalent to the vacuum solution of the Bianchi type-$VI_0$
spacetime. The second case allows non-vanishing massive and
nonlinear terms and depending on the sign of coupling constants
gives rise to expanding mode of expansion or the one that after
obtaining some maximum value contracts and ends in big crunch
generating spacetime singularity. This result once again shows the
sensitivity of spinor field to the gravitational one.

\vskip 0.1 cm

\noindent {\bf Acknowledgments}\\
This work is supported in part by a joint Romanian-LIT, JINR, Dubna
Research Project, theme no. 05-6-1119-2014/2016. I would also like
to thank the referee for some valuable suggestions that helped me to
improve the MS.

\end{document}